\documentstyle[12pt]{article}
\newcommand{\ka}{\kappa}
\newcommand{\la}{\lambda}
\newcommand{\Th}{\Theta}
\newcommand{\om}{\omega}
\newcommand{\be}{\begin{equation}}
\newcommand{\ee}{\end{equation}}
\newcommand{\beq}{\begin{eqnarray}}
\newcommand{\eeq}{\end{eqnarray}}
\newcommand{\lc}{\varepsilon}
\newcommand{\al}{\alpha}
\newcommand{\Ga}{\Gamma}
\newcommand{\si}{\sigma}
\newcommand{\fhi}{\varphi}
\newcommand{\bet}{\beta}

\begin{document}

\title{1-LOOP ANALYSIS OF THE PHOTON SELF-ENERGY DUE TO 3D-GRAVITY.}

\vspace{2cm}

\author{ C. Pinheiro, \\
Instituto de F\'\i sica e Qu\'\i mica - UFES,
Vit\'oria-ES, Brazil, \\
Instituto de F\'\i sica - UFRJ,
Rio de Janeiro-RJ, Brazil,
 \\ \\
G. O. Pires, \\
CBPF-DCP,
Rio de Janeiro-RJ, Brazil
\and  and \\
F. A. B. Rabelo de Carvalho, \\
Grupo de F\'\i sica Te\'orica - UCP,
Petr\'opolis-RJ, Brazil. \\ }

\footnotetext[1]{BitNet address   : Gentil@BRLNCC.BITNET }
\footnotetext[2]{InterNet address : Gentil@CBPFSU1.CAT.CBPF.BR }

\date{March, 1993.}

\maketitle

\vspace{2cm}

\begin{abstract}
{\it A Maxwell-Chern-Simons field is minimally coupled to
3D-gravity. Feynman
rules are written down and 1-loop corrections to the gauge-field self-energy
are calculated. Transversality is verified and gauge-field dynamical mass
generation does not take place.}
\end{abstract}

\pagebreak

\vspace{2cm}

The study of topologically massive Yang-Mills and gravity theories has raised a
great deal of interest after the classical paper by Deser, Jackiw and Templeton
\cite{DJT}. Relevant results, such as the finiteness of Chern-Simons theory
in Landau  gauge \cite{PS} and the renormalisability of 3D-gravity
\cite{DY,KK}, have contributed significantly for the broadening of research
on a number of aspects of D=3-gauge field theories.

\vspace{.3cm}

In a previous work \cite{NP}, we have addressed to the proposal of extending
the Barnes-Rivers projectors to account for 3D-gravity. As a byproduct of
our study, we propose to carry out perturbative calculations for Chern-Simons
gravity keeping the usual splitting of the metric field \cite{DY}.
For that, we perform the minimal coupling of a Maxwell-Chern-Simons field
to Einstein-Chern-Simons gravity and concentrate our attention on the
1-loop-corrected
Abelian gauge field self-energy. The main motivation behind our
calculation regards the analysis of the possibility of gauge-field dynamical
mass generation \cite{DJT,KS} through its gravitational coupling.

\vspace{.5cm}

Let us now begin by considering the Lagrangean we adopt to describe the
minimal coupling between the gauge and gravity sectors in D=3 :

\vspace{.3cm}

\be
{\cal L} = {\cal L}_{E.C.S.} + {\cal L}_{M.C.S.} \; ,
\ee

\vspace{.3cm}

\noindent where the first term on the right hand side denotes the
Einstein-Chern-Simons  Lagrangean,

\vspace{.3cm}

\be
{\cal L}_{E.C.S.} =  \frac{- 1}{2 \ka^{2}} \sqrt{-g} \; {\cal R} +
\frac{1}{\mu} \lc^{\la \mu \nu}
\Ga^{\rho}_{\;\;\la \si} ( \partial_{\mu} \; \Ga_{\rho \;\;\nu}^{\;\;\si} +
\frac{2}{3} \Ga_{\mu \;\;\fhi}^{\;\;\si} \; \Ga^{\;\;\fhi}_{\nu \;\;\rho} )
\; ,
\ee

\vspace{.3cm}

\noindent whereas the second term stands for the Maxwell-Chern-Simons term,
\footnote[3]{$ D_{\la} A_{\nu} \equiv \partial_{\la} A_{\nu} -
\Ga_{\nu \la}^{\rho} A_{\rho} $ is the covariant derivative under
gen. coord. transformations. }

\vspace{.3cm}

\be
{\cal L}_{M.C.S.} = \frac{- 1}{4} \sqrt{-g} \; g^{\mu \al} g^{\nu \bet}
F_{\mu \nu} F_{\al \bet} + \frac{M_{ph.}}{2}
\lc^{\mu \la \nu} A_{\mu} \; D_{\la} \; A_{\nu} \; .
\ee

\vspace{.5cm}

 Adopting the viewpoint of expanding the metric field around the
 flat-space geo\-metry, \footnote[4]{diag. $ \eta^{\mu \nu} \equiv (+\, ; - \,
, \cdots
 \, , \, -). $ }

\be
g^{\mu \nu}(x) = \eta^{\mu \nu} - \ka h^{\mu \nu}(x),
\ee

\vspace{.3cm}

\noindent where $ h^{\mu \nu} $ is the field variable defining the expansion,
and $ \ka $ is the Planck  constant, one can read off the propagators and
the Feynman rules describing the interaction in perturbative field theory.

\pagebreak

As fairly-well investigated \cite{DY,NP}, on the basis of the field
parametrisation (4), the propagator for topologically
massive gravity, in Feynman gauge, read as below :

\vspace{.3cm}

\beq
<\, h_{\mu \nu} (-q) \; h_{\ka \la}(q)\,> & = & \frac{-i}{ 64 \, q^{2} \,
 [q^{2} - M_{gr.}^{2}]  } \left\{
 32 \,i \, M_{gr.} q^{\al} \,
\left [ \lc_{\mu \al \la} \Th_{\ka \nu} + \lc_{\mu \al \ka} \Th_{\la \nu} +
\right. \right.  \nonumber \\  & + & \left.
 \lc_{\nu \al \la}\Th_{\ka \mu} + \lc_{\nu \al \ka}\Th_{\la \mu} \right ] +
\nonumber   \\  & - &
64  M_{gr.}^{2}
\left [ \eta_{\mu \ka} \eta_{\nu \la} + \eta_{\mu \la} \eta_{\nu \ka} -
2 \eta_{\mu \nu} \eta_{\ka \la} \right ] +
\nonumber \\ & - &
 64 q^{2} \left[ \eta_{\mu \ka} \om_{\nu \la} + \eta_{\mu \la} \om_{\nu \ka} +
\eta_{\nu \ka} \om_{\mu \la} +
\right.  \nonumber \\ & + & \left. \left.
 \eta_{\nu \la} \om_{\mu \ka} + \Th_{\mu \nu} \Th_{\ka \la} -
2 \eta_{\mu \nu} \om_{\ka \la} - 2 \eta_{\ka \la} \om_{\mu \nu} \right ] \,
\right \} ,
\eeq

\vspace{.3cm}

\noindent where $ \Th_{\mu \nu} $ and $ \om_{\mu \nu} $ are respectively
the usual transverse and longitudinal projectors in the space of vectors
and $ M_{gr.} \equiv ( \frac{\mu}{8 \ka^{2}} ) $.

\vspace{.5cm}

The well\--known Maxwell\--Chern\--Simons pho\-ton prop\-agator \cite{DJT}, in
Feyn\-man gau\-ge, take the form :

\vspace{.3cm}

\be
<\, A^{a} (-p) \; A^{b} (p) \,>  =  \frac{- i}{(p^{2} - M_{ph.}^{2})}
\left\{ \eta^{a b} - \frac{M_{ph.}}{p^{2}} \left[ M_{ph.}
\frac{p^{a} p^{b}}{p^{2}} - i \lc^{a \al b} \; p_{\al} \right] \right\} \; .
\ee

\vspace{.5cm}

As we are coupling a bosonic field to 3D-gravity, the affine connection,
$ \Ga_{\mu \;\; \rho}^{\;\;\nu} $, appearing in the gravitational
covariant derivative can be taken torsion-free; it is therefore identified
with the Christoffel symbol. So, the last term at the RHS of eq.(3)
shall not contribute to the interaction terms. Hence, the Feynman rules
required for the computation of the 1-loop photon self-energy can be
found by analysing the trilinear and quadrilinear parts stemming from
the Maxwell term minimally coupled to gravity :

\vspace{.3cm}

\be
{\cal L}^{(3)}_{M.} = \frac{- \ka}{4} \left( \frac{1}{2}
\eta^{\mu \al} \eta^{\nu \bet} h^{\fhi}_{\;\; \fhi}(x) -
\eta^{\mu \al} h^{\nu \bet}(x) - \eta^{\nu \bet} h^{\mu \al}(x) \right) \;
F_{\mu \nu} \; F_{\al \bet} \; ,
\ee

\vspace{.3cm}

\beq
{\cal L}^{(4)}_{M.} & = & \frac{- \ka^{2}}{4} \left( h^{\mu \al}(x) h^{\nu
\bet}(x) -
\frac{1}{2} \eta^{\mu \al} h^{\nu \bet}(x) h^{\fhi}_{\;\; \fhi}(x) + \right.
\nonumber \\ & - &
\frac{1}{2} \eta^{\nu \bet} h^{\mu \al}(x) h^{\fhi}_{\;\; \fhi} (x) -
\frac{1}{4} \eta^{\mu \al} \eta^{\nu \bet} h^{\fhi \si}(x) h_{\fhi \si}(x) +
\nonumber  \\ & + & \left.
\frac{1}{8} \eta^{\mu \al} \eta^{\nu \bet} h^{\fhi}_{\;\; \fhi}(x)
h^{\si}_{\;\; \si}(x)
\right) \;
F_{\mu \nu} \; F_{\al \bet} \; .
\eeq

\pagebreak

The 3- and 4-vertex Feynman rules can easily be read off from Lagrangean (3).
They are drawn in
Fig.(1) and their respective expressions, in momentum space, look as given
below :

\vspace{10cm}

\hspace{5.5cm} {\bf Fig.(1)}

\hspace{3.5cm}  {\it Graviton-photon vertices.}

\vspace{.3cm}

\beq
{\cal V}^{(3)}_{a b c d} & = & \frac{i \ka}{2} \left\{
(r \cdot p) \eta_{a b} \eta_{c d} - \eta_{a b} r_{c} p_{d} + \right.
\nonumber \\ & &
+ p_{a} r_{c} \eta_{b d} + 2 r_{b} p_{d} \eta_{a c} +
\nonumber \\ & & \left.
- 2 (r \cdot p) \eta_{a c} \eta_{b d} - 2 p_{a} r_{b} \eta_{c d} \right\}
\eeq

\noindent where $ p = q + r $,

\vspace{.3cm}

\noindent and

\vspace{.3cm}

\beq
{\cal V}^{(4)}_{a b c d e f} & = & i \ka^{2} \left\{
2 p_{c} r_{d} \eta_{a e} \eta_{b f} - 2 p_{c} r_{b} \eta_{a e} \eta_{d f} +
\right. \nonumber \\ & &
- (p \cdot r)\eta_{a e}\eta_{b f}\eta_{c d} + p_{f}r_{b} \eta_{a e} \eta_{c d}
+
\nonumber \\ & &
+ p_{a} r_{e} \eta_{b f} \eta_{c d} -  p_{a} r_{b} \eta_{c d} \eta_{e f} +
\nonumber \\ & &
- \frac{1}{2} (p \cdot r) \eta_{e f} \eta_{a c} \eta_{b d} +
\frac{1}{2} p_{f} r_{e} \eta_{a c} \eta_{b d} +
\nonumber \\ & & \left.
+\frac{1}{4} (p \cdot r) \eta_{e f} \eta_{a b} \eta_{c d}
- \frac{1}{4} p_{f} r_{e} \eta_{a b} \eta_{c d} \right\} \; ,
\eeq

\noindent where $ p + l = q + r $ .

\pagebreak

As long as the photon self-energy is concerned, there are only two
Feynman diagrams contributing at the 1-loop approximation. They are both
shown in \mbox{Fig.(2):}

\vspace{10cm}

\hspace{6cm} {\bf Fig.(2)}

\hspace{3cm}  {\it 1-loop contributions to photon self-energy.}

\vspace{.3cm}

\noindent The diagram presented in Fig.(2.a) yields the following
contributions to the self-energy :

\beq
{\cal I}^{(3)}_{e f}(p) & = & \int \frac{d^{3}\!q}{(2 \pi)^{3}} \;
{\cal V}^{(3)}_{a b e d}(p;q) \; <\, h^{a b}(-q) \; h^{c h}(q) \,> \;\times
\nonumber \\ & &
\times <\, A^{d} (-(p-q)) \; A^{g} (p-q) \,>  {\cal V}^{(3)}_{c h g f}(p;q) \;
{}.
\eeq

\vspace{.3cm}

\noindent The tadpole graph of Fig.(2.b) is given by the expression :

\be
{\cal I}^{(4)}_{e f}(p) = \int \frac{d^{3}\!q}{(2 \pi)^{3}} \;
{\cal V}^{(4)}_{a b c d e f}(p) \; <\, h^{a b}(-q) \; h^{c d}(q)\,>
 \; .
\ee

\vspace{.5cm}

Now, we simply replace in eqs.(11) and (12) the expressions previously
derived for the propagators and vertices.
The explicit evaluation of the loop integrals given above are algebrically
extremely laborious \footnote[5]{The algebraic manipulation of these loop
integrals and the task of exhautive simplifications would not be doable
without the use of the software 'FORM'. }. The $ {\cal I}^{(3)}-$ integral
generates 1512 terms, while the $ {\cal I}^{(4)}-$ integral other 140. A
careful analysis reveals that 54 \mbox{independent} momentum-space integrals
\footnote[6]{As this paper is intended to be a short letter, these Fynman
integrals shall be presented in a further work.} can be identifyed among
the generated terms. There appear integrals exhibiting up to 5 loop-momenta in
the numerator (for the sake of illustration, a representative
5-momentum integral is presented in the appendix).

\vspace{.3cm}

Dimensional regularisation procedure is adopted to solve the 1-loop integrals
\cite{DW}.
Clearly, since we are working in 3 dimensions, all these integrals turn out to
be finite. However, since the main motivation of this work concerns the
investigation of gauge-field dynamical mass generation through the
coupling to gravity, the evaluation of the Feynman integrals is an important
step in order to fix the answer for the graphs in terms of the external
momentum $ p^{\mu} $.

\vspace{.3cm}

The explicit calculations were carried out \cite{Next} and the final result
is that \underbar{no} topological mass term for the gauge field $ A^{\mu} $
appears at 1-loop upon the minimal coupling of this field to the gravitational
sector. This amounts to saying that the pole of the propagator given in
eq.(6) is not shifted after 1-loop corrections are taken into account.
Also, we should mention that the $ \cup (1)-$Ward identity is satisfied:
the total 1-loop contribuition to the gauge field self-energy diagram
has been checked to be transverse.

\vspace{.5cm}

Having understood that there is no dynamical mass generation for the
$ A^{\mu}-$field minimally coupled to gravity, we are now contemplating
interesting non-minimal couplings and we are trying to analyse
whether in these cases a gauge-field mass generation may take place.
These results shall soon be reported elsewhere \cite{Next}.

\vspace{.5cm}

We express our gratitude to the members of the
Theoretical Physics Group of the Universidade Cat\'olica de Petr\'opolis
for kind hospitality. We are also grateful to Adriana J. Simonato for kindly
drawing the figures of this work. G. O. Pires acknowledges Claudio A. Sasaki
for
suggestions with Maple and is indebted to Dr. Jos\'e A. Helay\"el-Neto
for invaluable discussions. CNPq and CAPES are acknowledged for the
financial support.

\pagebreak

\vspace{2cm}

\underbar{{\bf APPENDIX :}}

Typical 5-momentum integral :

\vspace{.5cm}

\beq
& &
 \int \frac{d^{3}\!q}{(2 \pi)^{3}} \;
\frac{p^{2}\;\; q^{\mu}q^{\nu}q^{\rho}q^{\si}q^{\la}}{
q^{4} (q^{2}-M_{gr.}^{2}) [ (p-q)^{2} - M_{ph.}^{2} ] (p-q)^{2} }\;\; =
\nonumber \\ &  & \nonumber \\ &  & =
\;\frac{p^{2}}{m_{gr.}^{2}}\;\left\{\;\frac{1}{m_{gr.}^{2}}\frac{1}{m_{ph.}^{2}}
\frac{i}{(4 \pi)^{\frac{3}{2}}} \sqrt{\pi} \int^{1}_{0} dx \,
\left[ \, - x^{3} ( {\cal A}_{1}^{\frac{1}{2}} - {\cal A}_{2}^{\frac{1}{2}} -
{\cal A}_{3}^{\frac{1}{2}} + {\cal A}_{4}^{\frac{1}{2}} ) \;
{\cal B}_{1}^{\mu \nu \rho \si \la}(p) \, +
\right. \right.
\nonumber \\  &  & \nonumber \\ &  & + \;\;
x^{5} ( {\cal A}_{1}^{-\frac{1}{2}} - {\cal A}_{2}^{-\frac{1}{2}} -
{\cal A}_{3}^{-\frac{1}{2}} + {\cal A}_{4}^{-\frac{1}{2}} )
\;\; {\cal B}_{2}^{\mu \nu \rho \si \la}(p) \;+
\nonumber \\ &  & \nonumber \\ & & + \;\; \left.
\frac{1}{3} x ( {\cal A}_{1}^{\frac{3}{2}} - {\cal A}_{2}^{\frac{3}{2}} -
{\cal A}_{3}^{\frac{3}{2}} + {\cal A}_{4}^{\frac{3}{2}} )
\;\; {\cal B}_{3}^{\mu \nu \rho \si \la}(p) \;\; \right] \; +
\nonumber \\ &  & \nonumber  \\ &  & - \;\;
\frac{1}{m_{ph.}^{2}}
\frac{i}{(4 \pi)^{\frac{3}{2}}} \sqrt{\pi} \int^{1}_{0}dx\;(1-x) \;\left[ \;
\frac{1}{2} x^{3} ( {\cal A}_{3}^{-\frac{1}{2}} - {\cal A}_{4}^{-\frac{1}{2}})
\;\; {\cal B}_{1}^{\mu \nu \rho \si \la}(p) \;\; \right. \; +
\nonumber \\ &  & \nonumber \\  &  & + \;\; \left. \left.
\frac{1}{2} x^{5} ( {\cal A}_{3}^{-\frac{3}{2}} - {\cal A}_{4}^{-\frac{3}{2}})
\;\; {\cal B}_{2}^{\mu \nu \rho \si \la}(p)
-\;\frac{1}{2} x ( {\cal A}_{3}^{\frac{1}{2}} - {\cal A}_{4}^{\frac{1}{2}})
\;\; {\cal B}_{3}^{\mu \nu \rho \si \la}(p) \;\;\right] \;\; \right\} \; ,
\nonumber
\eeq

\vspace{.5cm}

\noindent where the coefficients $ {\cal A}_{i} $ read as below :

\beq
{\cal A}_{1} & \equiv &
\left[ x(1-x) \cdot p^{2}-M_{ph.}^{2}\cdot x-M_{gr.}^{2}\cdot (1-x) \right] \;
,
\nonumber \\
{\cal A}_{2} & \equiv &
\left[ x(1-x) \cdot p^{2} - M_{gr.}^{2} \cdot (1-x) \right] \; ,
\nonumber \\
{\cal A}_{3} & \equiv &
\left[ x(1-x) \cdot p^{2} - M_{ph.}^{2} \cdot x \right] \; ,
\nonumber \\
{\cal A}_{4} & \equiv &
\left[ x(1-x) \cdot p^{2} \right] \; ,
\nonumber
\eeq

\vspace{.5cm}

\noindent and the tensors $ {\cal B}_{i}^{\mu \nu \rho \si \la}(p) $
take the forms :

\beq
{\cal B}_{1}^{\mu \nu \rho \si \la}(p) & \equiv & \left(
\eta^{\mu \la}p^{\nu}p^{\rho}p^{\si} + \eta^{\nu \la}p^{\mu}p^{\rho}p^{\si} +
\eta^{\rho \la}p^{\mu}p^{\nu}p^{\si} +
\right. \nonumber \\ & + &
\eta^{\si \la}p^{\mu}p^{\nu}p^{\rho} + \eta^{\mu \nu}p^{\rho}p^{\si}p^{\la} +
\eta^{\nu \si}p^{\mu}p^{\rho}p^{\la} +
\nonumber \\ & + &
\eta^{\rho \si}p^{\mu}p^{\nu}p^{\la} + \eta^{\mu \rho}p^{\nu}p^{\si}p^{\la} +
\eta^{\nu \rho}p^{\mu}p^{\si}p^{\la} +
\nonumber \\ & + & \left.
\eta^{\mu \si}p^{\rho}p^{\nu}p^{\la}
\right) \; ,
\nonumber \\
{\cal B}_{2}^{\mu \nu \rho \si \la}(p) & \equiv & \left(
p^{\mu}p^{\nu}p^{\rho}p^{\si}p^{\la} \right) \; ,
\nonumber \\
{\cal B}_{3}^{\mu \nu \rho \si \la}(p) & \equiv & \left(
\eta^{\mu \nu}\eta^{\rho \la}p^{\si} + \eta^{\mu \nu}\eta^{\si \la}p^{\rho} +
\eta^{\nu \si}\eta^{\mu \la}p^{\rho} +
\right. \nonumber \\ & + &
\eta^{\nu \si}\eta^{\rho \la}p^{\mu} + \eta^{\rho \si}\eta^{\mu \la}p^{\nu} +
\eta^{\rho \si}\eta^{\nu \la}p^{\mu} +
\nonumber \\ & + &
\eta^{\mu \rho}\eta^{\nu \la}p^{\si} + \eta^{\mu \rho}\eta^{\si \la}p^{\nu} +
\eta^{\nu \rho}\eta^{\mu \la}p^{\si} +
\nonumber \\ & + &
\eta^{\nu \rho}\eta^{\si \la}p^{\mu} + \eta^{\mu \si}\eta^{\rho \la}p^{\nu} +
\eta^{\mu \si}\eta^{\nu \la}p^{\rho} +
\nonumber \\ & + & \left.
\eta^{\mu \nu}\eta^{\rho \si}p^{\la} + \eta^{\nu \rho}\eta^{\mu \si}p^{\la} +
\eta^{\mu \rho}\eta^{\nu \si}p^{\la}
\right)\; . \nonumber
\eeq

\end{document}